# *Automatedfirst-order design of double-sided -sided telecentric zoom systems based on PSO algorithms*

ZICHAO FAN, SHILI WEI, ZHENGBO ZHU, YAN MO, YIMING YAN, AND DONGLIN MA*

*School of Optical and Electronic Information, Huazhong University of Science and Technology, Wuhan, Hubei 430074, China*
*Corresponding author: madonglin@hust.edu.cn*

**In this letter, we propose an efficient and robust method to retrieve an optimal initial configuration for the design of double-sided telecentric zoom system by Particle Swarm Optimization (PSO) algorithm. We demonstrate that the proposed algorithm is much more efficient than Monte Carlo in designing zoom systems with two fixed foci as well as a given zoom ratio. Furthermore, a compact initial design of three-component 4X zoom system with two fixed foci is proposed to show the high efficiency and great potential of our proposed algorithm in searching proper initial configuration for complex optical systems.**

Telecentric zoom optical systems [1-2] have important applications in many areas such as machine vision and industrial metrology. It is well worth to find an appropriate starting point for the zoom system design. Feasible starting points are very difficult to be acquired for complex systems with advanced specifications (zoom ratio, total length, and zoom curves). The starting point that is far from the prescribed design may cost a large amount of time on optimization or result in a worse design [3]. Moreover, a plausible starting point with a small solvable region might lead to an impracticable system after optimization. Consequently, retrieving a reasonable starting point for complex systems is labor-intensive and highly relies on designers' experience.

An optimal and feasible starting point configuration is especially important for the final design of double-sided telecentric zoom systems. The design of the double-sided telecentric zoom systems are introduced in [2, 4-6], and they all provided a strict analytical derivation of paraxial designs for double-sided telecentric zoom lens systems. Their work played a pioneering role in designing double-sided telecentric systems and established a basic analytical method. In particular, the configuration proposed by Miks solves the problem that Kryszczyński's solution is not flexible enough [2, 4]. However, a simple and practical way to find a suitable starting point for double-sided telecentric zoom system is not presented as far as we know.

In this letter, we propose a novel design method to search an optimal starting point of the double-sided telecentric zoom system automatically. It is friendly for less experienced designers and greatly increases efficiency. To perform a complex design with the algorithm, only the knowledge of evaluation performance metric as well as some particular constraints is required.

We start the study by employing an initial theoretical derivation based on paraxial analysis proposed by Miks [6]. We will use the three-element double-sided telecentric zoom lens as an example to introduce our proposed algorithm and analyze the design results in detail. It should be noted that our proposed design algorithms can be applied to find an initial configuration for any other complex optical system [5].

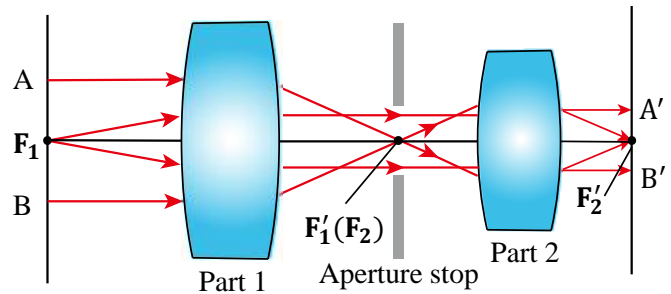

Fig. 1. Configuration of double-sided telecentric zoom lens.

The principal structure of double-sided telecentric zoom lens is shown in Fig. 1. The first part of the system is a fixed focal system with a focal length of $f_{p1}$ ($f_{p1}> 0$). The second part is a zoom focal system with a continuously varying focal length of $f_{p2}$ within a certain range of $[f_0, tf_0]$. The rear focal point $\mathbf{F}'_1$ of part 1 and the front focal point $\mathbf{F}_2$ of part 2 are commonly located at the center of the aperture stop. The major task in the design of a double-sided zoom system is to search a good starting point of the zoom part under current physical limitations.

We concentrate on finding the configuration with a larger zoom range and a shorter total length. Related calculation is based on the

Gaussian bracket method, which is specifically introduced in Ref. [6-8]. Based on Ref. [7], the first-order parameters of the optical system can be expressed as:

$$\begin{cases} \varphi = -\gamma \\ S_F = \delta/\gamma \\ S'_F = -\alpha/\gamma \end{cases}, \quad (1)$$

where $\varphi$ is the optical power of the zoom system, $S_F$ is the front focal distance (distance of the front focal point from the first lens), $S'_F$ is the back focal distance (distance of the rear focal point from the last lens), and the definitions of all other parameters can be found in Ref. [7]. If we specify $d_1$ as the distance of the second component from the first component, and $d_2$ as the distance of the third component from the second one, then the distance of the rear focus from the front focus $D$ can be speculated by:

$$-S_F + d_1 + d_2 + S'_F = D. \quad (2)$$

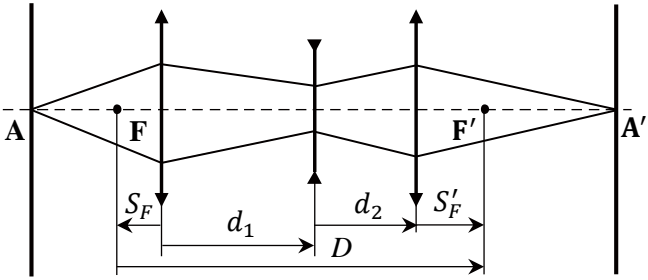

Fig. 2. Paraxial configuration of the three-element zoom system with fixed distance between the front focus and the rear focus.

We still need another constraint to find the solution of the equations. Considering Petzval sum, we can have

$$\varphi_1 + \varphi_2 + \varphi_3 = 0. \quad (3)$$

where $\varphi_1$, $\varphi_2$, $\varphi_3$ are the optical power of the three components respectively. Combining Eqs. (1)-(3), we can obtain as follows [8]:

$$a_3 d_1^3 + a_2 d_1^2 + a_3 d_1 + a_4 = 0, \quad (4)$$

where $a_i = a_i(\varphi_1, \varphi_2, \varphi, D)$, and $d_2$ can be calculated by:

$$(\varphi_1 + \varphi_2)(\varphi_1 + \varphi_2 - \varphi_1\varphi_2 d_1)d_2 + (\varphi_1^2 d_1 - \varphi) = 0. \quad (5)$$

We find that Eq. (4) can be reformulated in the following terms:

$$A(d_1 - \frac{\varphi_1 + \varphi_2}{\varphi_1 \varphi_2})(b_2 d_1^2 + b_1 d_1 + b_0) = 0, \quad (6)$$

where $A = A(\varphi_1, \varphi_2)$ and $b_i = b_i(\varphi_1, \varphi_2, \varphi, D)$.

We can firstly calculate $d_1$ by solving Eq. (6), and then derive $d_2$ by Eq. (5). Looking back to the form of Eq. (6), we can directly assume that $d_1 = (\varphi_1 + \varphi_2)/\varphi_1\varphi_2$. Substituting the expression of $d_1$ into Eq. (5), we can find that the coefficient of $d_2$ is equal to 0, and then the expression of $d_1$ can be simplified as $d_1 = \varphi/\varphi_1^2$. The value of $d_1$ cannot satisfy simultaneously these two relations with the focal length varying from $f_0$ to $tf_0$. As a result, finding the solution of Eq. (6) is equivalent to solving the following quadratic equation:

$$b_2 d_1^2 + b_1 d_1 + b_0 = 0 \quad (7)$$

The values of parameters $\varphi_1, \varphi_2, D, f_0$ can be optimized to find a proper starting point in our proposed algorithm by employing the thought of Particle Swarm Optimization (PSO) algorithm [9-12]. Each particle searches for a better position in the search space by changing its velocity according to rules originally inspired by behavioral models of bird flocking. In the algorithm, a whole set of candidate solutions to the optical design can be taken as a swarm, each element in the swarm can be regarded as a particle, and the number of particles $N$ is called the population size. The proposed algorithm can help to retrieve the initial structure of the system, as well as to optimize the system's refined structure [11].

During the initialization process, the initial state of each particle in the population is randomly assigned. Each particle in the swarm is in an $S$-dimensional vector space and its position can be regarded as $\mathbf{x}_i^k = [\varphi_1, \varphi_2, D, f_0]^T \in \mathcal{X}$ ($S$=4) and $\mathcal{X}$ is the boundary of search space. The position of a particle represents a candidate solution to the optimization problem of the optical system. We estimate a most probable interval for each variable of the initial state inside the boundary for $\mathbf{x}_i^k$. The direction of flocking is determined by the personal best position, the global best position and inertia. Corresponding velocity of the particle can be expressed as a vector $\mathbf{v}_i^k = [v_1, v_2, v_D, v_f]^T$. The subscript represents the sequence of a particle as $i$ ($i$=1, 2, ..., $N$), and the superscript represents the $k$-th iteration. During the optimization process, we employ the fitness function $T=T(D, S_F)$ in PSO algorithms as the merit function to find a more compact configuration. The merit function can be defined as follows:

$$T = \begin{cases} w_T \cdot (\max(|S_F|) - L_0) + D & \max(|S_F|) \geq L_0 \\ D, & \max(|S_F|) < L_0 \end{cases}, \quad (8)$$

where $w_T$ is the penalty factor for designs with the maximum front focal distance larger than given length $L_0$ during zooming.

Unlike the fixed focal system, the initial state of our system consists of a group of multiple configurations with the focal length varying from $f_0$ to $tf_0$. In the traditional PSO algorithm with simply evaluating the performance of a single configuration, we cannot avoid the collision problem between neighboring components in the motion process.

In order to overcome the issue, we pioneer the use of another module to evaluate the zoom process during the update phase. To escape the collision issue, we divide the whole zoom range into $M$ intervals and check the practicability of each configuration with focal length of $f = f_0 + j \cdot (t-1) \cdot f_0 / M$ ($j$=1,2,..., $M$+1) respectively. Obviously, the root of the Eq. (6) continuously varies with $f$, which means that we can always find the solution with no collision for the whole zooming process when $M$ is large enough.

Fig. 3 shows that the particles are screened and evaluated, and then $\mathbf{v}_i^k$ is determined by the sum of the three weighted vectors in our proposed algorithm. To prevent being misled by some impractical particles with lower merit function values, particles that satisfy the physical limitations are selected and the best position will be determined by merit function values among them. Then all particles adjust their directions with reference to their reliable peers and their own right experience in the update. What needs to be noted is that $\varphi_1$, $\varphi_2$, and $D$ are fixed when analyzing multiple configurations of each particle with its focal length varying from $f_0$ to $tf_0$. Particles update their states based on the following rules:

$$\mathbf{v}_i^k = w \cdot \mathbf{v}_i^{k-1} + w_1 \cdot ran \cdot (\mathbf{p}_{best}^k - \mathbf{x}_i^k) + w_2 \cdot ran \cdot (\mathbf{g}_{best} - \mathbf{x}_i^k),$$
$$\mathbf{x}_i^{k+1} = \mathbf{x}_i^k + \mathbf{v}_i^k, \tag{9}$$

where $w$ is the inertia coefficient, $\mathbf{p}_{best}$ is the personal best position of the particle in history, $\mathbf{g}_{best}$ is the global best position among the group of particles in the current iteration, $w_1$ and $w_2$ are acceleration coefficients, and $ran$ is a random value ranging from 0 to 1.

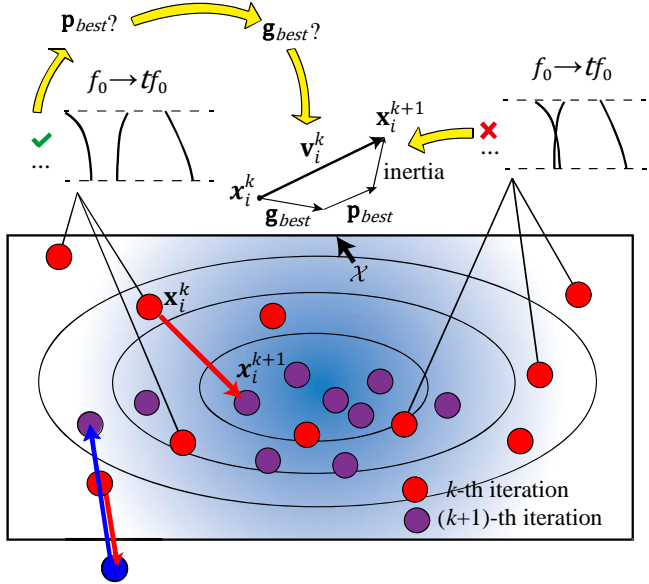

Fig. 3. The movement of particles in our algorithm.

Unavoidably, there may be cases where particles escape the boundary $\mathcal{X}$. Therefore, we add a vector in the direction opposite to the updated direction for these particles:

$$\mathbf{x}_i^{k+1} = \mathbf{x}_i^k - w_b \mathbf{v}_i^k, \tag{10}$$

where $w_b$ is the return coefficient ($w_b \geq 1$), which is to adjust the magnitude of this vector. The blue clip in Fig. 3 shows that these particles reverse their directions and retire back to the boundary. The general process of modified PSO algorithm is depicted in Fig. 4.

Firstly, we apply the proposed algorithm to design a 2X zoom system with fixed distance between two foci to verify the feasibility. Given a random initial state, we obtain an optimized starting point (x=[41.14 -10.80 37.04 35.08]) by our algorithm by setting $w_T = 0.25$ and $L_0 = 10$. In order to visually present the optimization process, we reduce the dimensions to a two-dimension space with $f_1$ and $f_2$ fixed ($f_1 = 41.14$ and $f_2 = -10.80$) and run the algorithm again to verify the optimized results of $D$ and $f_0$. Fig. 5(a)-(c) intuitively demonstrate the process of convergence and Fig. 5(d) provides the final convergence process of our given merit function during optimization. We have checked that the particles on the right side of Fig. 5(a) have lower merit function values but fall outside the specified boundary, which however is still acceptable. In Fig. 5(b)-(c), the particles that do not satisfy the physical limitations will not affect the directions of the group and finally almost all particles converge to neighborhood of the global best position. As is shown in Fig. 5(d), the merit function can converge quickly when optimized by the PSO algorithm, which indicates that the PSO algorithm can be very efficient for the search of optimal values and will not fall into the local optimal values or unsolvable region, and therefore is potentially beneficial to optical design. Similarly, any other two variables can converge to the initially acquired optimal position by our PSO algorithm.

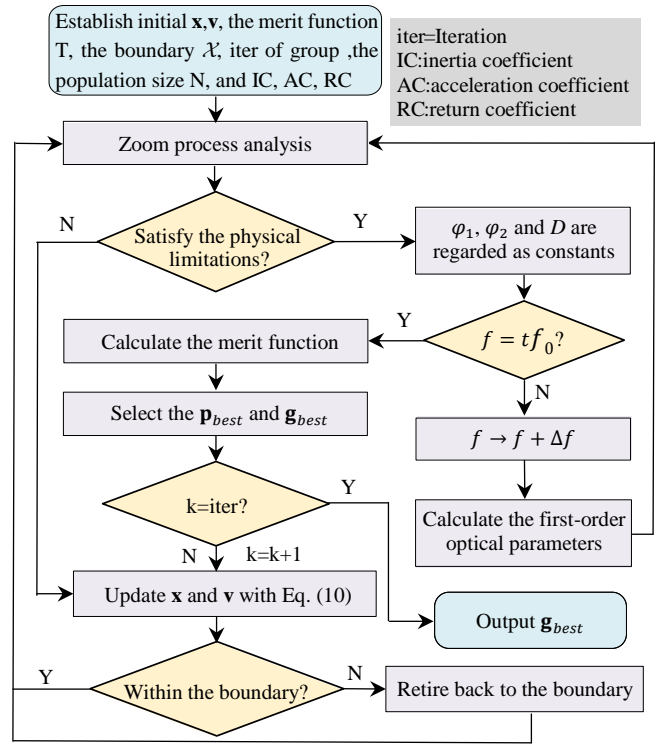

Fig. 4. Flow chart of modified PSO algorithm.

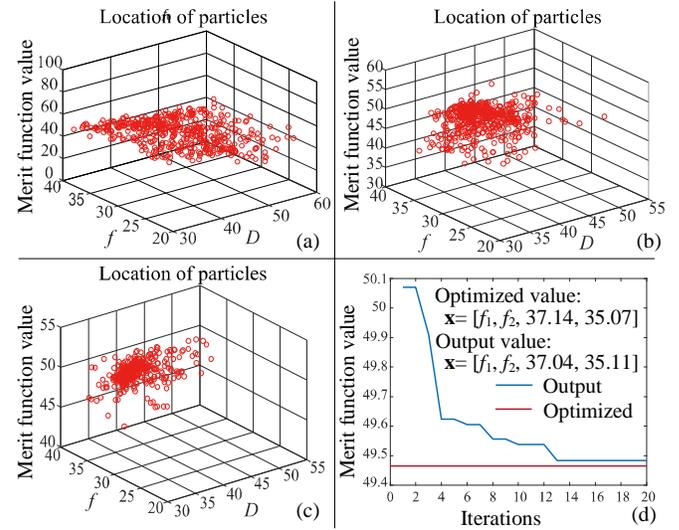

Fig. 5. Convergence process of optimization iterations: (a) initial state with iter=1; (b) intermediate state with iter =10; (c) optimized state with iter =20; (d) convergence process of merit function.

Then we compare PSO algorithm with Monte Carlo simulation by exploring the possible zoom curves in different zoom ratios.

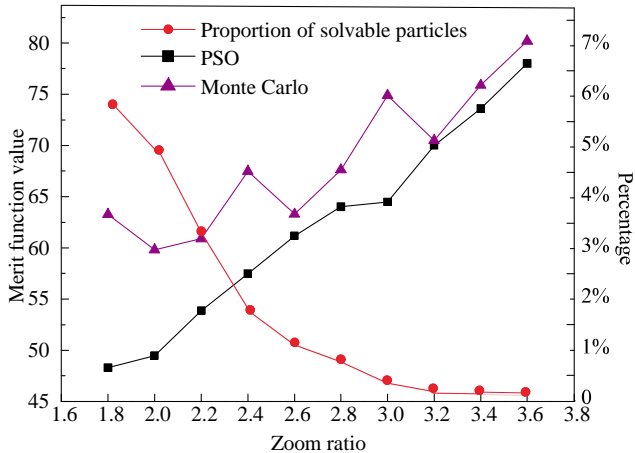

Fig. 6. Optimized merit function value vs. zoom ratio under different algorithms and proportion of particles in solvable region vs. zoom ratio.

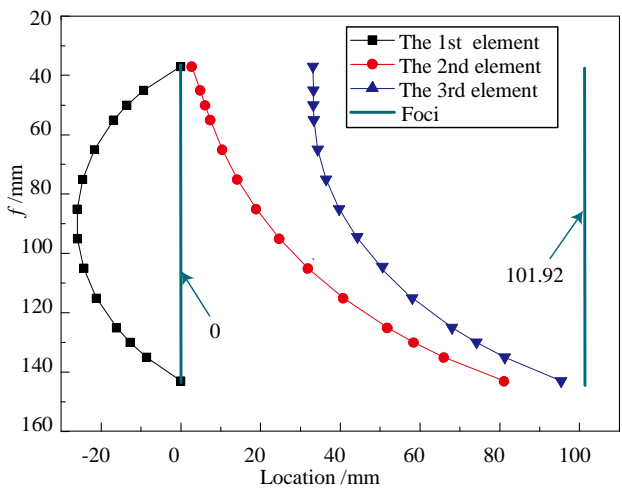

Fig. 7. Optimized results: loci of the three elements and the two foci of the 4X zoom system during zooming.

Optimal design results derived by the two algorithms are recorded in Fig. 6. Obviously, the best starting point obtained by PSO is much better than that by Monte Carlo, since PSO can lead to a relatively lower merit function value for any designed zoom ratio. Furthermore, it takes several hours to run Monte Carlo algorithm, which is dozens of times longer than the PSO algorithm. In the meanwhile, we put 10,000 particles into a wide solution space for testing, and then verifies the probability of finding a feasible starting point without rich experience. As shown in Fig. 6, the solvable region becomes very narrow at high zoom ratio, where it is almost impossible for us to find a solution by trial and error approach. However, some few particles within or nearby the solvable region can lead their peers to an advisable position in PSO algorithm, which perfectly solves the needle-in-a-haystack problem of finding a feasible starting point for double-sided telecentric zoom systems at high zoom ratio.

Moreover, we apply the algorithm to design the zoom system with a zoom ratio of 4 as well as a fixed distance between the two foci. The detailed design result is shown in Table 1. Fig. 7 shows the optimized locus of each component as a function of $f$, as well as the loci of the front and rear foci. As can be seen from the design results, our zoom range has been increased from 2X to 4X compared to the data given previously, and the distance between the two foci has been reduced by about 1/3, which is much better than previous designs [2]. Meanwhile, the optimized design enjoys a smooth zoom trajectory, motionless fixed foci, and continuous zoom. Besides the optimal solution, any one of a series of other good similar results can be chosen as an alternative starting point for the final optimization, which provides great convenience to optical designers.

**Table 1. Parameters of three-element zoom lens (mm)**

| $f_1 = 92.71$ | | $f_2 = -24.37$ | | $D = 101.92$ |
|---|---|---|---|---|
| $f$ | $d_1$ | $d_2$ | $S_F$ | $S'_F$ |
| 37 | 2.74 | 30.37 | 0.00 | 68.55 |
| 85 | 44.89 | 20.80 | 25.99 | 62.22 |
| 143 | 81.09 | 14.30 | 0.00 | 6.384 |

Traditional trial and error approach for designing a zoom system with two fixed foci highly relies on the guidance, experience, and skill of designers, such as the choice of starting points, computer aided optimization, and so on. Therefore, it is inefficient, high-threshold, and sometimes even does not perform very well. In this letter, we propose a novel optimization method based on modified PSO algorithm to search an initial configuration for the double-sided telecentric zoom system. The algorithm can greatly help release the dependence on designer's experience and extensively save the human effort in the design work. We also demonstrate that our proposed modified PSO algorithm is very efficient in searching the global optimal starting point. With the algorithm, a series of good solutions can be automatically retrieved, and designers can pick the best one among them according to actual requirements. To summarize, retrieving a starting point for designing complex optical systems based on advanced algorithms is very important to fulfilling the automatic optical design, as well as the application of artificial intelligence in optics design in the future.

**Funding.** Wuhan Municipal Science and Technology Bureau (2017010201010110); Huazhong University of Science and Technology (2017KFYXJJ026); Natural Science Foundation of Hubei Province (2018CFB146); National Natural Science Foundation of China (61805088).